\documentclass[conference]{IEEEtran}

\usepackage{amsmath}
\usepackage[tight]{subfigure}
\usepackage[footnotesize]{caption} 
\usepackage{cite}
\usepackage{eurosym}
\usepackage{amsfonts}
\usepackage{graphics}
\usepackage{epsfig}
\usepackage{psfrag}
\usepackage{pstricks}
\usepackage{array}
\usepackage{shortvrb}
\usepackage{epsf}
\usepackage{graphicx}
\usepackage{rotating}
\usepackage{float}
\usepackage{color}
\usepackage{cite}
\usepackage{soul}
\usepackage{multirow}
\usepackage{acronym}
\usepackage{url}
\usepackage{breakurl}
\usepackage[breaklinks, hidelinks]{hyperref}

\acrodef{ess}[\textsc{ess}]{Energy Storage System}
\acrodef{dae}[\textsc{dae}]{Differential Algebraic Equation}
\acrodef{vsc}[\textsc{vsc}]{Voltage Sourced Converter}
\acrodef{coi}[\textsc{coi}]{Centre of Inertia}
\acrodef{fdf}[\textsc{fdf}]{Frequency Divider Formula}
\acrodef{wecs}[\textsc{wecs}]{Wind Energy Conversion System}
\acrodef{spvg}[\textsc{spvg}]{Solar Photo-Voltaic Generation}
\acrodef{tcl}[\textsc{tcl}]{Thermostatically Controlled Load}
\acrodef{hvdc}[\textsc{hvdc}]{High-Voltage Direct Current}
\acrodef{pll}[\textsc{pll}]{Phase-Locked Loop}
\acrodef{pmu}[\textsc{pmu}]{Phasor Measurement Unit}
\acrodef{rtds}[\textsc{rtds}]{Real-Time Digital Simulator}
\acrodef{emt}[\textsc{emt}]{Electromagnetic Transients}
\acrodef{tg}[\textsc{tg}]{Turbine Governor}
\acrodef{avr}[\textsc{avr}]{Automatic Voltage Regulator}
\acrodef{agc}[\textsc{agc}]{Automatic Generation Control}
\acrodef{gps}[\textsc{gps}]{Global Positioning Satellite}

\DeclareGraphicsExtensions{.eps,.png,.pdf}

\def \R {{\rm I\kern -2.2pt R\hskip 1pt}}

\newcommand{\PreserveBackslash}[1]{\let\temp=\\#1\let\\=\temp}

\IEEEoverridecommandlockouts
\begin{document}

\title{Stability Assessment of Low-Inertia Power Systems: A System Operator Perspective}

\author{ \IEEEauthorblockN{Manuel Hurtado,\IEEEauthorrefmark{1} \IEEEmembership{IEEE~Member}, Mohammad Jafarian,\IEEEauthorrefmark{1} Taulant K\"{e}r\c{c}i,\IEEEauthorrefmark{1}
    \IEEEmembership{IEEE~Member},  
Simon Tweed,\IEEEauthorrefmark{1} \\Marta Val Escudero,\IEEEauthorrefmark{1} Eoin Kennedy,\IEEEauthorrefmark{1}
    and
    Federico~Milano,\IEEEauthorrefmark{2}~\IEEEmembership{IEEE~Fellow}}\vspace*{0.3cm}
  \IEEEauthorblockA{
    \begin{tabular}{cc}
      \begin{tabular}{@{}c@{}}
        \IEEEauthorrefmark{1}
        Transmission System Operator\\
        Innovation \& Planning office, EirGrid, plc\\
        Ireland
      \end{tabular} &
      \hspace{0.3cm}
      \begin{tabular}{@{}c@{}}
        \IEEEauthorrefmark{2}
        School of Electrical and Electronic Engineering \\
        University College Dublin \\
        Ireland
      \end{tabular} 
    \end{tabular}
  }
  %
  %
}

\IEEEoverridecommandlockouts

\maketitle
\IEEEpubidadjcol

\begin{abstract}
  This paper discusses the stability assessment of low-inertia power systems through a real-world large-scale low-inertia system, namely, the All-Island power system (AIPS) of Ireland and Northern Ireland.  This system currently accommodates world-record levels of system non-synchronous penetration namely 75\% (planning to increase to 80\% next year).  The paper discusses one-month results obtained with the state-of-the-art stability tool called look-ahead security assessment (LSAT).  This tool carries out rotor-angle, frequency and voltage stability analyses and is implemented in the control centres of the transmission system operators (TSOs).  The paper shows that, at the time of writing, the main binding stability constraint of the AIPS is related to the limits on the rate of change of frequency (RoCoF).    
\end{abstract}


\section{Introduction}
\label{sec:intro}

\subsection{Motivation}
\label{Motivation}

Large-scale low-inertia power systems are characterised by high penetration of inverter-based resources (IBRs) such as wind and solar photovoltaic (PV).  The power system community have recently updated the classification of power system stability to account for the changing dynamic behavior mainly due to the introduction of such technologies \cite{9286772}.  However, the dynamics of such systems are still to be fully studied and understood \cite{8450880}.  In particular, in the literature, there is no final conclusion on what is the main stability constraint that limits the penetration of non-synchronous generation in low-inertia power systems.  This paper attempts to answer this question by running dynamic stability studies on a real-world large-scale low-inertia power system, namely the All-Island power system (AIPS) of Ireland (IE) and Northern Ireland (NI) which allows up to 75\% of system non-synchronous penetration (SNSP) \cite{10253224}.

\vspace{-2mm}
\subsection{Literature Review}
\label{sec:literature}

The topic of low-inertia power systems and, in particular, technical (dynamic) constraints associated with it has been growing fast in recent years\cite{RATNAM2020109773, HONG20211057, AHMED2023486, gfm}.  For example, the authors in \cite{9361257} study the  small-signal stability issue in low-inertia systems and show that depending on the power system network and the mix of generators considered (i.e., synchronous, grid-following, grid-forming), among others, one can obtain completely different  penetration levels of IBR units (e.g., 60\% or 93\%).  However, small and simplified benchmark systems were used and the focus was only on small-signal stability.  The authors in \cite{9804725} study the large-signal (transient) stability problem in low-inertia systems in terms of the coupling between the electromagnetic dynamics of the IBRs and the electro-mechanical dynamics of the synchronous generators.  But again no real-world power system model is considered to validate the study results and the focus is only on one stability problem.  In \cite{CHEN2020106475},  the hypothetical scenario of 100\% IBRs using an AIPS model is studied.  However, the focus is only on frequency stability and the AIPS model is based on guessed dynamic data of the devices.  Similarly, the authors in \cite{6805233} study the frequency stability challenges in the AIPS based on a futuristic model of the AIPS and do not consider other stability problems.  
\begin{figure}[t!]
  \begin{center}
    \resizebox{1.0\linewidth}{!}{\includegraphics{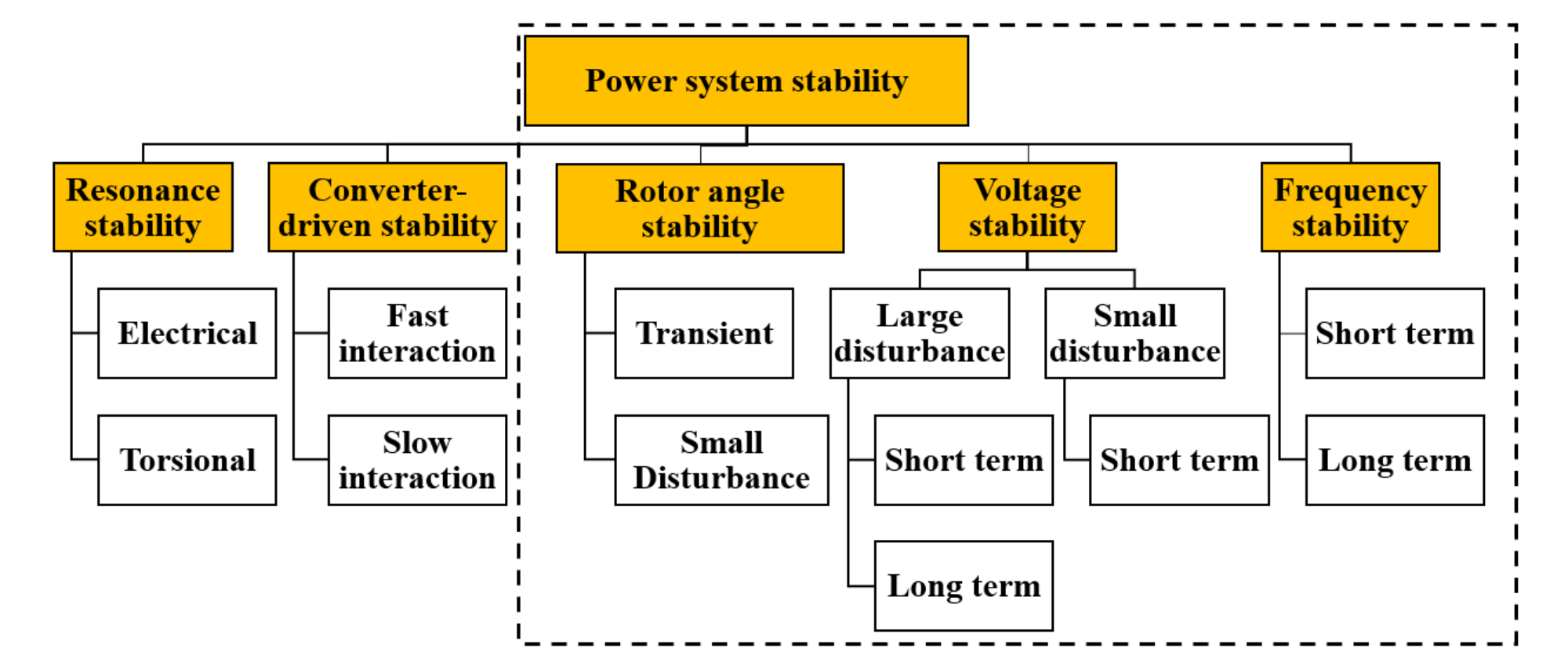}}
    \caption{Classification of power system stability \cite{9286772}.}
    \label{fig:class}
  \end{center}
  \vspace*{-0.3cm}
\end{figure}

This paper addresses the above limitations by assessing transient, frequency and voltage stability in low-inertia systems using a real-world, fully-fledged dynamic power system model, namely the AIPS (see Fig.~\ref{fig:class}).  The dynamic performance of the grid is evaluated by means of the state-of-the-art tool look-ahead security assessment (LSAT) implemented in the control centres of EirGrid and SONI, the transmission system operators (TSOs) of IE and NI, respectively \cite{esig}.   With regard to the new stability phenomena in Fig.~\ref{fig:class}, the AIPS has so far experienced a few cases of very-low frequency oscillations and sub-synchronous torsional interactions.  However, these phenomenons are identified post-events as currently the TSOs do not have the real-time capability to evaluate them pre-event.  Due to the nature of the AIPS, the TSOs expect potential risks associated with these phenomenons.  The TSOs are working toward having the study capabilities (e.g., electromagnetic transients) to analyze them in the future and, for this reason, these emerging phenomena are outside the scope of this paper \cite{eirgrid}.

\subsection{Contributions}
\label{Contributions}

The specific contributions of this paper are the following:
\begin{itemize}
\item An analysis of conventional stability phenomena based on a real-world low-inertia system, namely the AIPS. 
\item Demonstrate through the analysis that, currently, the main stability problem is related to frequency stability, and, in particular, to high RoCoF.
\end{itemize}

\subsection{Paper Organization}

The paper is structured as follows.  Section \ref{sec:background} provides a background on how the TSOs manage the dynamic stability in the AIPS.  In particular, this section describes the main operational constraints in place to ensure system stability and provides an overview of LSAT.  Next, Section \ref{sec:case} presents and discusses the results of the dynamic studies based on June 2023 and discusses the relationship between different variables of the systems.  Finally, Section \ref{sec:conclu} draws the main findings of the case study.

\section{Dynamic Stability Management in the Ireland and Northern Ireland Power Systems}
\label{sec:background}

EirGrid and SONI manage dynamic stability in the operational time frames using different dynamic operational constraints \cite{10253224}, and by performing online dynamic assessment through LSAT.  With regard to the operational constraints, the TSOs have in place four operational constraints/limits namely \cite{eirgrid}: (i) an SNSP limit; (ii) a minimum number of conventional units online (MUON); (iii) a rate of change of frequency (RoCoF) limit; and (iv) a minimum inertia floor.  The TSOs are also planning to introduce a new dynamic stability metric called ``System Strength'' to maintain stable operation of IBRs during high SNSP scenarios, among others \cite{eirgrid}.  The current and expected evolution of these constraints are shown in Table \ref{tab:constraint}. In particular, it is worth mentioning that RoCoF $\pm1$ Hz/s became an enduring operational constraint in May 2023.  Because of this, we provide below a brief description of the rationale and practical issues behind the RoCoF limit change (from $\pm0.5$ Hz/s to $\pm1$ Hz/s).  Similarly, an overview of the real-time component of LSAT which is used to evaluate the dynamic stability in the AIPS is provided in Section \ref{sec:lsat} below.   
\begin{table}[h!]
  \centering
  \caption{Evolution of operational policy constraints in the AIPS \cite{eirgrid}.} 
  \label{tab:constraint}
  \begin{tabular}{cccccc}
    \hline
    Year & SNSP & RoCoF & Inertia & MUON & System Strength \\
    \hline
    2023  & 75\% & $\pm 1$ Hz/s & 23 GWs & 7 & Under development \\
    2030 & 95\% & $\pm 1$ Hz/s & 20 GWs & 3 & Enduring policy \\
    \hline
  \end{tabular}
\end{table}
%

\subsection{RoCoF Setting Change}
\label{sec:rocof}

More than a decade ago an ambitious target of 40\% of electricity to originate from renewable generation by 2020 was set, most of this from wind turbines.  The 2010 published “Facilitation of Renewables” (FoR) study identified that having 40\% of IBRs would have a significant impact on RoCoF experienced in the network.  For this reason, it was essential to increase the maximum RoCoF from $\pm 0.5$ to $\pm 1$ Hz/s.  The FoR study also identified that during times of high wind generation and following the loss of the single largest credible contingency, RoCoF values greater than $\pm 0.5$ Hz/s but no greater than $\pm 1$ Hz/s could be experienced.

Following the conclusions of the FoR, generator market participants and distribution system operators worked towards changing the RoCoF setting on all assets to $\pm1$ Hz/s. This change involved real field testing of generators to showcase the withstand capability.  After running RoCoF compliance tests for all large conventional generators, a RoCoF trial started in June 2020.  It was envisaged the trial project would broadly comprise detailed technical studies and simulations to identify potential RoCoF-related vulnerabilities in the system and a trial of operating the power system with an increased RoCoF limit.  The expectation was following successful completion of the trial, and subject to appropriate mitigation strategies being established, the RoCoF operational policy range would be widened from the threshold of $\pm 0.5$ Hz/s to $\pm 1$ Hz/s on an enduring basis.  In May 2023, the RoCoF trial was concluded and approved (following a successful trial and detailed studies), and is now allowing operating the AIPS with a RoCoF limit of $\pm 1$ Hz/s.  This enduring limit is enabling higher levels of renewable energy in the AIPS.

\begin{figure}[h!]
  \begin{center}
    \resizebox{1.0\linewidth}{!}{\includegraphics{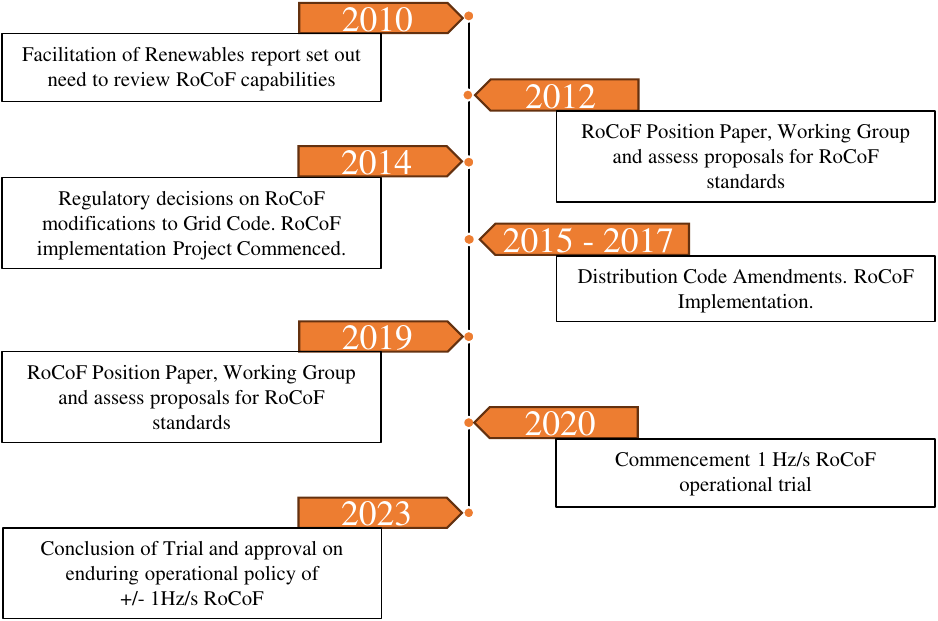}}
    \caption{Roadmap of RoCoF settings change in the AIPS.}
    \label{fig:ro}
  \end{center}
  \vspace*{-0.3cm}
\end{figure}

In summary, the whole process of changing the RoCoF took more than a decade.  The roadmap of RoCoF settings change in the AIPS is summarized in Fig.~\ref{fig:ro}. 


\subsection{Look-ahead Security Assessment Tool (LSAT)}
\label{sec:lsat}

%
\begin{figure}[t!]
  \begin{center}
    \resizebox{0.97\linewidth}{!}{\includegraphics{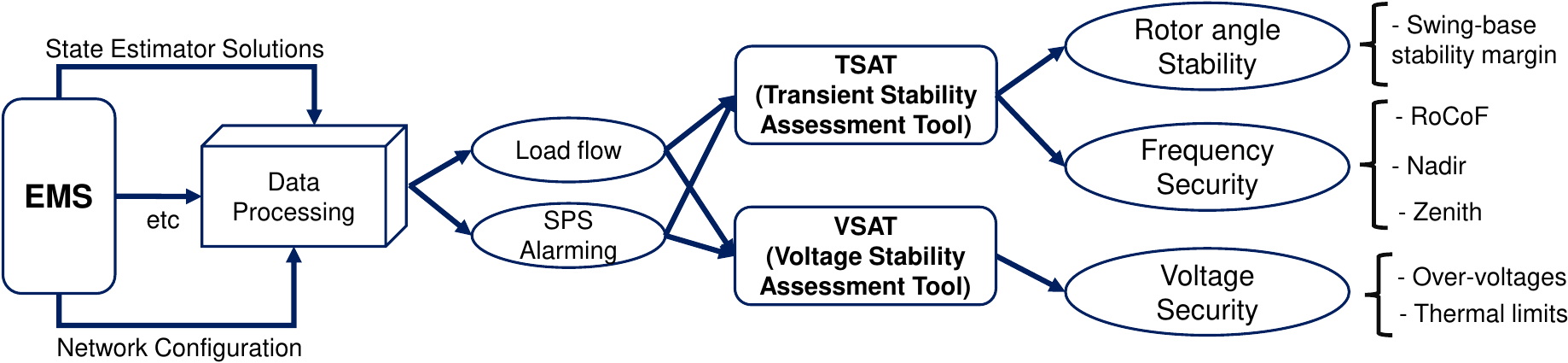}}
    \caption{Overview of real-time LSAT.}
    \label{fig:lsat}
  \end{center}
  \vspace*{-0.3cm}
\end{figure}

Figure~\ref{fig:lsat} provides an overview of the real-time component of LSAT.  It went live the TSOs control centres in 2010 and has been operating since then to asses the security of the AIPS in terms of rotor angle stability, frequency security, and voltage security \cite{esig}.  Note that security is a broader term which includes stability \cite{10105838}.  For instance, a system may be stable but not necessarily secure (e.g., stable frequency but outside operational limits), but not the other way around.  From December 2020, the look-ahead functionality was integrated into LSAT with the aim of assessing the security of the network in the near future, e.g., 10 hours ahead.  Given that the focus of this paper is on real-time stability problems, the look-ahead functionality results of LSAT are out of subject and not considered here.  LSAT is composed of two major components:

\paragraph{Transient Security Assessment Tool (TSAT)} This tool conducts dynamic analyses to assess the rotor angle stability, based on the swing-base stability margin criterion \cite{Kundur:94}, and frequency security, based on Nadir, Zenith, and RoCoF criteria. 
 In particular, RoCoF is calculated using a rolling 500 ms period and filtering is applied to eliminate high frequency
transients (e.g., during transmission faults).  
\paragraph{Voltage Security Assessment Tool (VSAT)} This tool conducts power flow analysis to assess the voltage security in quasi steady state, based on over-voltage and thermal limit criteria.  The assessment is carried out both for real-time and near future (look-ahead) time horizons.  The focus of this paper is on real-time assessment of the network security. Regarding real-time assessment, in every 5 minutes, LSAT receives the network configuration and state estimator solutions, together with other data, from the energy management system (EMS). This data is processed and fed into VSAT and TSAT.  These components, in parallel, run around 800 N-1 contingencies in less than 5 minutes and notify about insecure contingencies. 

LSAT serves as a critical decision-support tool for control room operators and provides radar-like guidance on how to operate the power system in a safe and secure manner.  The interested reader is referred to \cite{esig} for additional information on LSAT and its use in the control centres of the TSOs.

\section{Case Study}
\label{sec:case}

To asses the stability of the AIPS, we use the real-time LSAT analysis and results for one relevant month, in this case, June 2023.  This month was particularly of interest as the network experienced lower levels of inertia.  However, our study shows that the trend extracted from the results of this month is consistent to that in previous months. 

We evaluate each stability problem based on certain predefined
metrics/criteria.  These are presented in Table \ref{tab:criteria}.
In particular, for security reasons and to account for potential
uncertainties such as in modelling and data accuracy, the TSOs have
build up different margins for different stability metrics.  For
example, a RoCoF of $\pm 0.9$ Hz is used instead of the operational
limit of $\pm 1$ Hz/s.  Similarly, the Zenith limit set in LSAT is
50.8 Hz instead of 51 Hz operational limit. Therefore, all the results
presented are with respect to these margins.  On the other hand,
rotor-angle stability is evaluated by means of the negative margin
metric which captures the maximum rotor angle difference between any
pair of generators across the network \cite{margin}.  Note that if any
of the limits are reached during the simulation, it will be
highlighted to the operators that we are at the security limits, but
the system is not unstable per se.  Control room operators will take
then the necessary mitigating actions in a timely manner to address
such cases.

\begin{table}[t!]
  \centering
  \caption{Rotor-angle, voltage and frequency security limits in LSAT.} 
  \label{tab:criteria}
  \begin{tabular}{cccccc}
    \hline
    Binding constraint & Security criteria 
     \\
    \hline
    Rotor-angle  & Negative margin \cite{margin}   \\
    Voltage & Outside Grid Code ranges \cite{code1} \\
    RoCoF  & $\pm \rm$0.9 Hz/s \\ 
    Zenith  & 50.8 Hz \\ 
    Nadir  & 49.0 Hz \\ 
    \hline
  \end{tabular}
\end{table}

\subsection{Overview of Transient, Frequency and Voltage Insecurities}
\label{sec:stochastic}

The total number of LSAT cases run for June 2023 is 8594.  As mentioned above, each case evaluates around 800 N-1 contingencies. Table~\ref{tab:insecure} presents all the relevant statistics for the three stability problems namely rotor-angle, voltage and frequency.  First, it is interesting to see that out of 8594 cases only 418 cases (or approximately 4.86\%) are reported as insecure.  
Note that in all those cases, the system was secure in the basecase where no contingency was applied to the network.  This number of insecurities is to be expected as the TSOs push the operational boundaries of the AIPS.  In fact, the number of insecure cases is minimal considering the original stability margins set by the TSOs (e.g., $\pm0.9$ Hz/s instead of $\pm1$ Hz/s).  It is also worth mentioning that power systems cannot operate 100\% securely as it is cost-prohibitive if at all physically feasible \cite{10105838}.  While comparing the \% of the three different stability problems, it can be seen that frequency is the main problem with 2.22\% of total insecure cases.  In particular, it appears that RoCoF is the main problem with 1.35\% of total cases.  However, note that in practice frequency stability has significantly improved in the AIPS in terms of Nadir and Zenith in recent years \cite{10253411}. 

\begin{table}[t!]
  \centering
  \caption{Summary of total rotor-angle, frequency and voltage cases where the constraint is binding for June 2023.} 
  \label{tab:insecure}
  \begin{tabular}{cccccc}
    \hline
    Binding constraint & Total cases & \% of all cases & Comparative \%  
     \\
    \hline
    Rotor-angle  & 67 & 0.78\% & 16.03\% \\
    Voltage & 160 & 1.86\% & 38.28\% \\
    RoCoF  & 116 & 1.35\%  & 27.75\%\\ 
    Zenith  & 49 & 0.57\% & 11.72\%  \\ 
    Nadir  & 26 & 0.30\% & 6.22\%  \\ 
    \hline
  \end{tabular}
\end{table}

Among those security issues, our investigation reveals that all
rotor-angle stability issues were local e.g, due to two units
oscillating against the whole system, and hence, they were not related
to the system being low in inertia.  Similarly, the voltage
insecurities were mainly concerned with not having enough resources to
absorb extra reactive power generation in a certain part of the system
and then again, not related to the inertia of the system.  In other
words, both of these insecurity issues are not related to low-inertia
scenarios.  Because of this, and since frequency stability appears the
dominant stability problem for this particular month (more than 45\%
of the total insecure cases), we analyze in more detail each frequency
stability metric and in which system conditions they appear more in
the next section.

\subsection{Correlation between Frequency Insecurity and Operating Conditions }
\label{sec:correlation}

In this section, we are interested to identify the system-wide contributing factors (specific variables of the system) to insecurities in order to have a better understating of the trends that lead to those insecurities.

\subsubsection{Inertia}
\label{sec:rocof1} 

\begin{figure}[t!]
  \begin{center}
    \resizebox{1.0\linewidth}{!}{\includegraphics{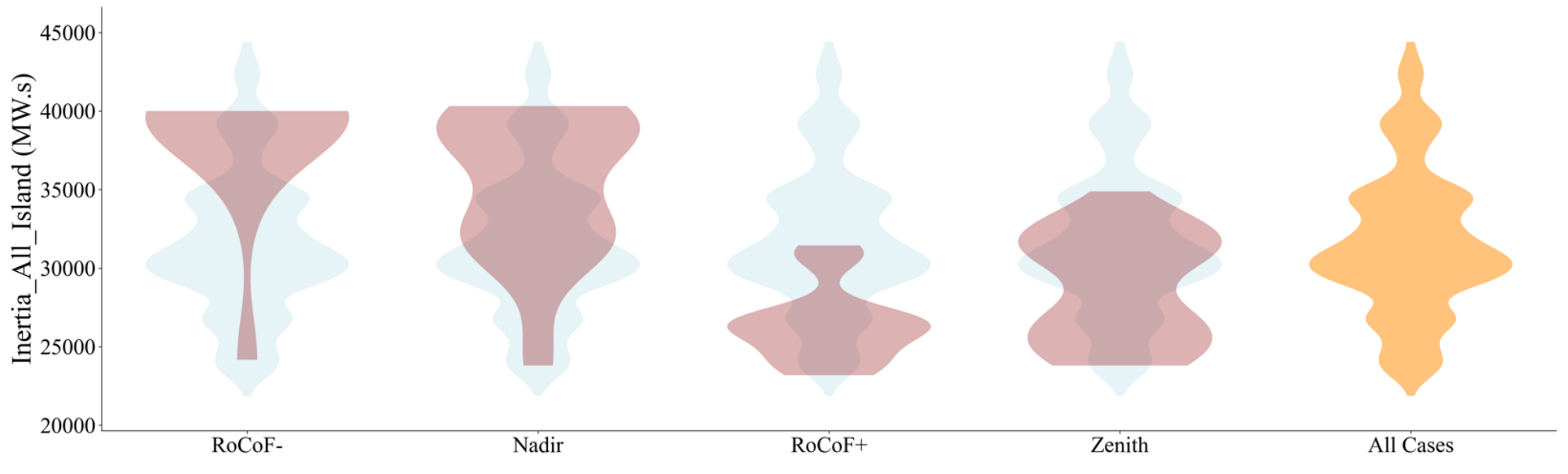}}
    \caption{\color{blue} Correlation between inertia and frequency insecurities.}
    \label{fig:inertia}
  \end{center}
  \vspace*{-0.3cm}
\end{figure}

Figure~\ref{fig:inertia} shows the correlation between total inertia
in the AIPS and RoCoF-, Nadir, RoCoF+, Zenith and all cases,
respectively.  Blue colour means secure cases while red represent
insecure cases.  There is a strong correlation between RoCoF- and
Nadir and high All-Island inertia.  While this may be
counter-intuitive, the system conditions in specific areas of the AIPS
following a contingency can be such that support this correlation.
For example, it is well-known that in case of the loss of the
North-South Tie-line between IE and NI (system separation) and
depending on the infeeds/outfeeds in IE and NI, frequency stability
issues (e.g., RoCoF- and Nadir) can appear \cite{eirgrid}.  In fact,
this is one of the main reasons why the TSOs have a requirement for a
minimum number of large conventional units online in both
jurisdictions \cite{10253224}.  To further support this, the TSOs have
recently identified through detailed dynamic studies that there is a
need to procure 4 GWs of inertia in NI (incentivizing one zone within
NI) and 6 GWs of inertia in IE (incentivizing two zones within IE)
\cite{synchcond}.  On the other hand, as expected, there is strong
correlation between low inertia and RoCoF+ and Zenith.  The figure
also shows that there is, overall, a medium correlation between
inertia and all cases (including both secure and insecure).

\subsubsection{Demand}
\label{sec:demand}

Figure~\ref{fig:demand} shows the results of the correlation between demand and all frequency insecurities.  Similar to above, it can be seen that there is a strong correlation between high demand and RoCoF- and Nadir.  On the other hand, low demand scenarios appear to lead to RoCoF+ and Zenith issues.  For example, during night hours with a lot of wind generation
and low demand can lead to such potential insecurities.  Figure~\ref{fig:demand} also shows that, overall, demand has a strong impact in all cases.


\subsubsection{Wind}
\label{sec:wind}

\begin{figure}[t!]
  \begin{center}
    \resizebox{1.0\linewidth}{!}{\includegraphics{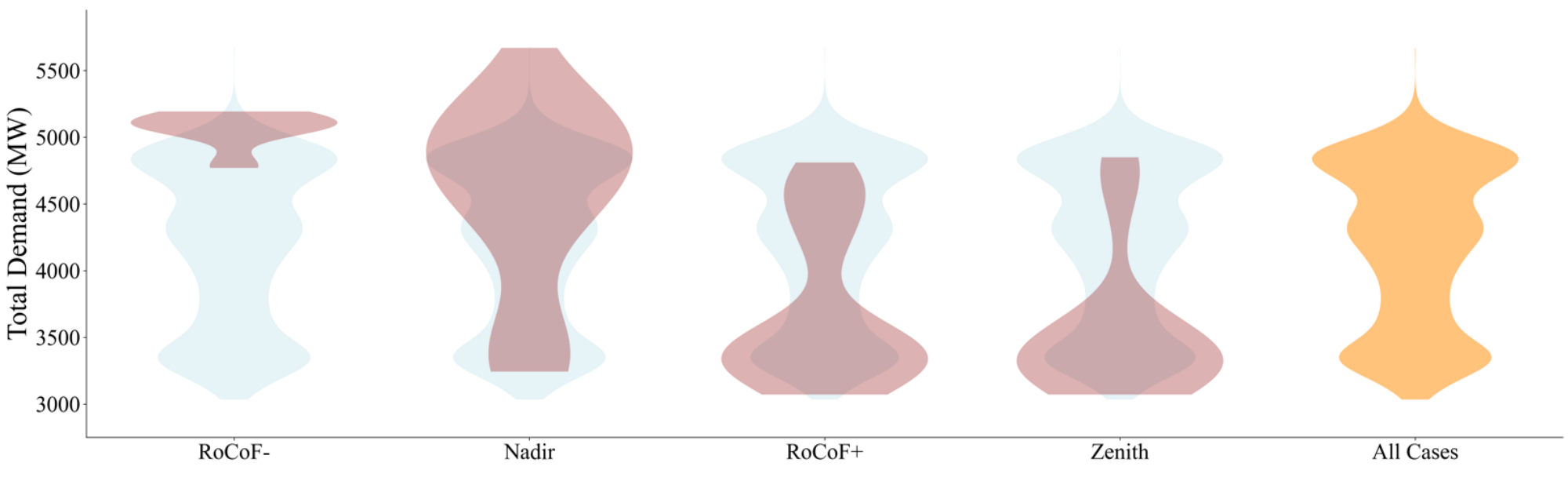}}
    \caption{Correlation between total demand and frequency insecurities.}
    \label{fig:demand}
  \end{center}
  \vspace*{-0.3cm}
\end{figure}

\begin{figure}[t!]
  \begin{center}
    \resizebox{1.0\linewidth}{!}{\includegraphics{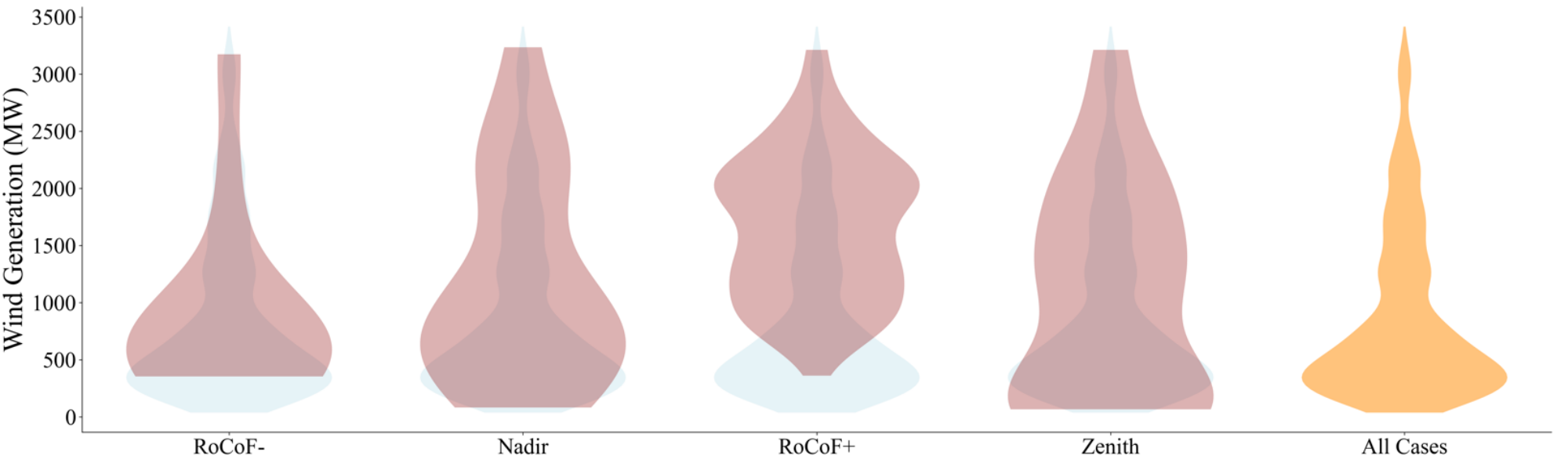}}
    \caption{Correlation between wind generation and frequency insecurities.}
    \label{fig:wind}
  \end{center}
  \vspace*{-0.3cm}
\end{figure}

The correlation results between wind and all frequency insecurities are shown in Figure~\ref{fig:wind}.  In general, wind appears to have the strongest correlation with the frequency insecurities compared to inertia (Fig.~\ref{fig:inertia}) and demand (Fig.~\ref{fig:demand}).  Specifically, it appears that in the vast majority of cases low wind leads to RoCoF- and Nadir insecurities.  Similar to the inertia explanation above, these insecurities are driven by specific network conditions following N-1 contingencies.  On the contrary, relatively medium and high wind scenarios appear to lead to potential Zenith and RoCoF+ insecurities.


\subsection{Frequency Stability}
\label{sec:freq}

To have a better understanding of the relationship between different system conditions and frequency insecurities, we plot the above variables (inertia, demand, and wind) against each other and all LSAT cases (including the insecure ones) by means of scatter plots.

\subsubsection{Frequency Nadir}
\label{sec:nadir}

Figure~\ref{fig:nadir} shows different scatter plots for frequency Nadir and Zenith.  Specifically, the total LSAT cases including the insecure cases are plotted as a function of: (i) demand vs wind generation; (ii) inertia vs wind generation; and (iii) demand vs inertia.  It is interesting to observe that frequency Nadir happens generally during periods of high demand, low wind and relatively high inertia.  While these results seem somewhat counter-intuitive, they can be explained by the fact that in high demand scenarios and low wind, all the generators are at high outputs and the interconnectors are also importing, thus, leading to potential extreme Nadirs.

\subsubsection{Frequency Zenith}
\label{sec:zenith}

\begin{figure}[t!]
  \begin{center}
    \resizebox{1.0\linewidth}{!}{\includegraphics{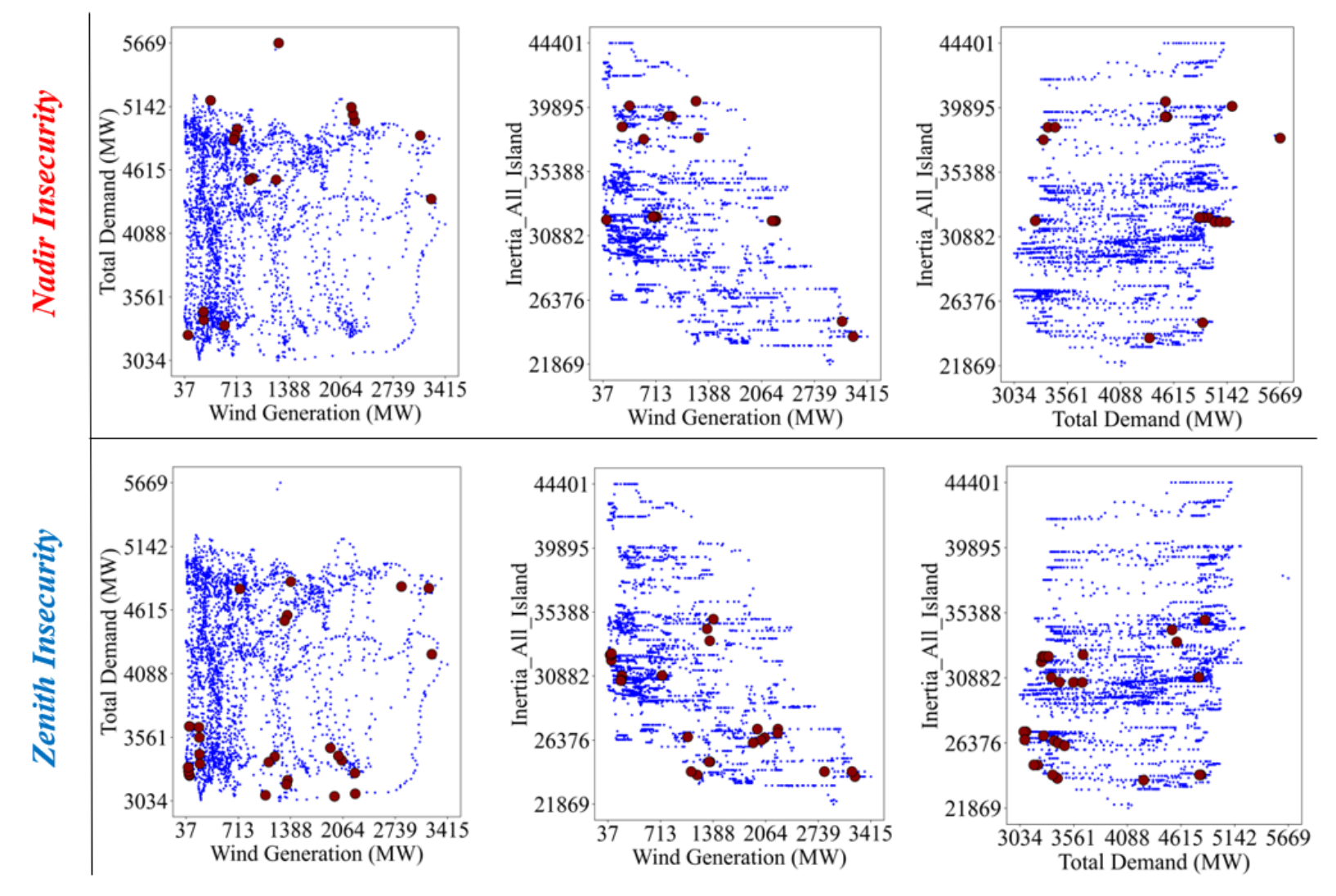}}
    \caption{Relationship between different system conditions (inertia, demand, and wind) and Zenith and Nadir insecurities.}
    \label{fig:nadir}
  \end{center}
  \vspace*{-0.3cm}
\end{figure}

Regarding frequency Zenith constraint, Fig.~\ref{fig:nadir} shows that those usually happen during periods of low demand and relatively high wind generation and low inertia (e.g., overnight).  As opposed to Nadir results discussed above, these are somewhat expected as, for example, during nights with high wind generation (and also low demand/inertia) tripping of interconnectors exporting lead to over-frequency. 

\subsubsection{RoCoF+ and RoCoF-}
\label{sec:rocof2}

\begin{figure}[t!]
  \begin{center}
    \resizebox{1.0\linewidth}{!}{\includegraphics{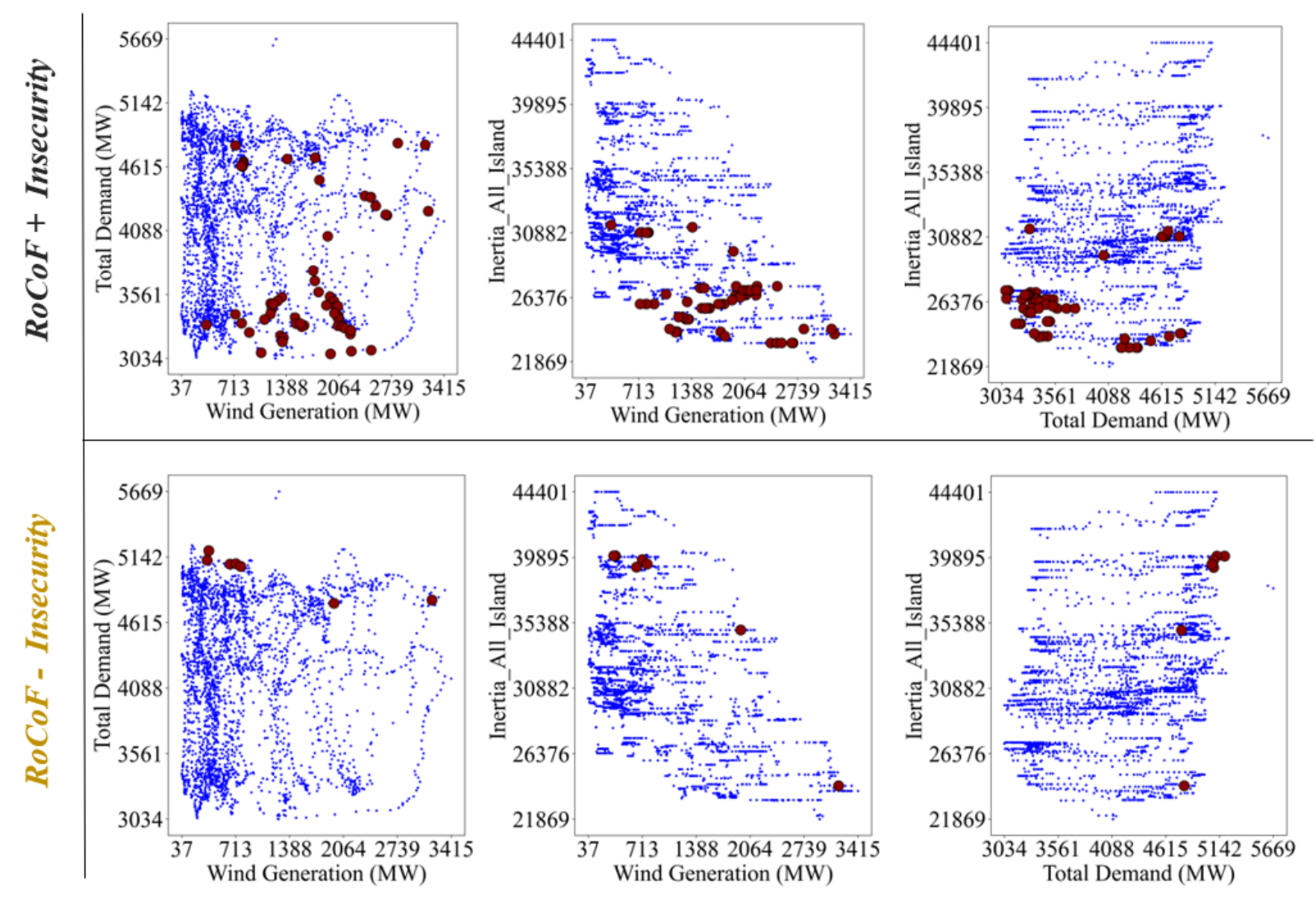}}
    \caption{Relationship between different system conditions (inertia, demand, and wind) and RoCoF insecurities.}
    \label{fig:rocof}
  \end{center}
  \vspace*{-0.3cm}
\end{figure}

Figure~\ref{fig:rocof} shows the results for RoCoF including both positive and negative values.  RoCoF+ insecurities happen predominately during periods of high wind and low inertia and demand scenarios.  Similar to Zenith results, this is expected as a light power system with high non-synchronous penetration (wind) is prone to frequency related problems in case of large-scale contingencies.  Figure \ref{fig:rocof} also shows that RoCoF- insecurities occur more often during periods of low wind, and high demand and inertia scenarios (similar to Nadir results).


\section{Conclusions}
\label{sec:conclu}

This paper deals with the real-time stability assessment of
low-inertia power systems.  The focus is on conventional stability
problems namely rotor-angle, frequency and voltage stability.  To do
so, we use a real-world large-scale low-inertia system namely the AIPS
that currently accommodates world-record levels of system
non-synchronous penetration namely 75\% (planning to increase to 80\%
next year).  Using the state-of-the-art stability tool LSAT, and its
stability results for one month, namely June 2023, the study reveals
that, at the time of writing, the main binding constraint in the AIPS
is related to frequency stability (e.g., due to low-inertia), and in
particular, the limits on the RoCoF.  Note that we are aware that this
conclusion might differ for other large-scale low-inertia systems
(e.g., can have other binding dynamic constraints such as system
strength) \cite{8779818}.  As part of Shaping Our Electricity Future
(SOEF) Roadmap, the TSOs are procuring low-carbon inertia services and
plan to perform a product review of all reserve products to address
future system needs (e.g., high RoCoF) \cite{soef}.



\end{document}